\documentclass[a4paper,prd,superscriptaddress,twocolumn,aps,notitlepage,floatfix,showpacs]{revtex4}
\usepackage[usenames]{color}
\usepackage[pdftex, unicode, colorlinks, bookmarks=true, citecolor=blue, link color=blue, urlcolor=blue]{hyperref}
\usepackage{amsmath, amssymb, amsfonts}
\usepackage{graphicx}
\graphicspath{{../PI_in_aLIGO_materials/}}

\newcommand{\SD}[1]{{#1}}

\begin{document}
\title{Time evolution of parametric instability in large-scale gravitational-wave interferometers.}

\author{Stefan L.~Danilishin}
\email{stefan.danilishin@ligo.org}
\affiliation{UWA, School of Physics, 35 Stirling Hwy, Crawley, WA 6009, Australia}

\author{Sergey P. Vyatchanin}
\affiliation{M.V.Lomonosov Moscow State University, Faculty of Physics, Moscow 119991, Russia}
\author{David G. Blair}
\author{Ju Li}
\author{Chunnong Zhao}

\affiliation{UWA, School of Physics, 35 Stirling Hwy, Crawley, WA 6009, Australia}


\begin{abstract}
We present a study of three-mode parametric instability in large-scale gravitational-wave detectors. Previous work used a linearised model to study the onset of instability. This paper presents a non-linear study of this phenomenon, which shows that the initial stage of exponential rise of the amplitudes of a higher order optical mode and the mechanical internal mode of the mirror is followed by a saturation phase, in which all three participating modes reach a new equilibrium state with constant oscillation amplitudes. Results suggest that stable operation of interferometers may be possible in the presence of such instabilities, thereby simplifying the task of suppression.
\end{abstract}
\maketitle

\begin{figure*}[ht]
 \includegraphics[width=\textwidth]{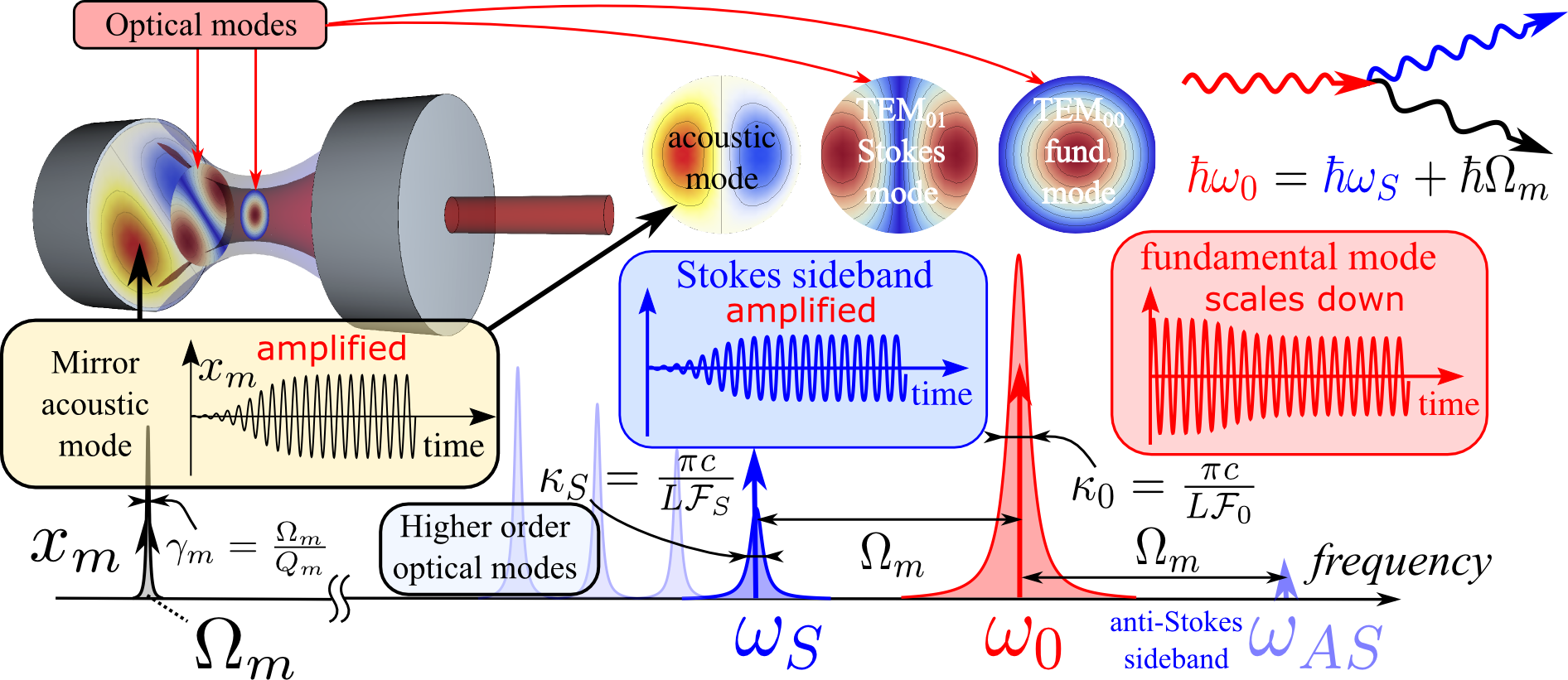}
 \caption{Schematic of the 3-mode interaction that gives rise to parametric instability in optomechanical systems: (i) acoustic modes of the mirror, excited by thermal fluctuations, create motional sidebands of the pump mode carrier frequency offset by acoustic frequency $\Omega_m$, (ii) the lower frequency Stokes sideband is enhanced by a higher-order optical cavity mode. A beat note between this and the fundamental pump mode creates a radiation pressure force at the acoustic frequency $\Omega_m$, (iii) this leads to a growth of the acoustic oscillation amplitude, which, in turn, increases the amplitude of the Stokes sideband, thereby raising radiation pressure force amplitude and closing the instability feedback loop. The strength of the 3-mode optomechanical interaction, and therefore the chance of instability development, depends on the following three factors: (i) The extent to which the spatial distributions of all 3 participating modes overlap, characterised by overlapping factor $\Lambda_{0S}$, defined in Eq.~\eqref{eq:Overlap_def}, (ii) The accuracy of 3-mode frequency tuning -  \textit{i.e.} detuning must be smaller than the larger of optical modes bandwidth, $\Delta_m = \omega_0 - \omega_S - \Omega_m \ll \max[\kappa_0,\,\kappa_S]$, and (iii) The energy loss rates in all 3 modes, which need to be lower than the rate of power transfer between the modes, which itself is characterised by the coupling strength $G_{0S}$. These 3 conditions yield the definition of parametric gain, $\mathcal{R}_0$, given in Eq.~\eqref{eq:PIcrit_res}, and the condition for parametric instability (PI) given by Eq.~\eqref{eq:PIcrit_det}.}\label{fig:PI_dynamics}
\end{figure*}  

\section{Introduction}
Three mode parametric instability in large scale, high optical power gravitational wave detectors was predicted by Braginsky \textit{et al.} in 2001 \cite{2001_Braginsky_PI_paper}. All subsequent analyses \cite{Vyatchanin:2012,Strigin200710,2009_CQG_Gras_et_al,2010_PLA_Evans_et_al} relied on the model prediction where amplitudes of certain acoustic modes of the interferometer mirrors would grow exponentially once an instability threshold of input laser power is reached. It was generally considered that the exponential growth would eventually render the whole setup unstable and cause an interferometer to lose lock. 

This prognosis, however, relied on a linearised approximation of the 3-mode optomechanical interaction which is valid only for small amplitudes of acoustic and Stokes modes. For larger values, it is intuitively obvious that non-linearity should ultimately modify this growth. If the optical configuration can be maintained one would expect that the acoustic and higher-order optical oscillations should saturate. Knowing the amplitudes of such saturation effects as well as their timescale is of crucial importance for the operation of the real detectors now being implemented. 

It is hard to overestimate the significance of rigorous analysis of this phenomenon in large-scale gravitational wave interferometers. Second generation detectors, such as Advanced LIGO \cite{Thorne2000,Fritschel2002} are at the the latest stages of construction and testing. These instruments are planned to have up to $\sim800$~kW of circulating laser power in the arms. As demonstrated in \cite{Strigin200710, 2007_Quant.Electronics.37.1097_Strigin_Vyatchanin,2007_Phys.Lett.A.362.91_Gurkovsky,2009_CQG_Gras_et_al,2010_PLA_Evans_et_al}, the chance of 3-mode parametric instability at such high level of power is very high. Similarly, other advanced detectors,  Advanced Virgo \cite{Acernese2006-2}, KAGRA \cite{KAGRA_paper_Somiya} and GEO-HF\cite{Willke2006}, might be susceptible to this effect, though with different probability (see review \cite{Vyatchanin:2012} for details). 

Therefore, knowledge of the temporal dynamics and the values of final amplitudes the three participating modes reach at the saturation stage allows to design a feedback control system to suppress this instability before it develops. Several methods of mitigation this phenomenon were developed that can benefit from this information: (i) varying of the mirrors radii of curvature by heating \cite{2005_PhysRevLett.94.121102_Zhao_et_al,2010_CQG.27.205019_Gras_et_al,2007_JOSAB.24.1336_Degallaix}; (ii) decreasing acoustic modes Q-factor \cite{2009_CQG.26.015002_Ju_Li_et_al,2009_CQG.26.135012_Gras_et_al,2010_CQG.27.205019_Gras_et_al}; (iii) introducing additional damping to acoustic modes via electrostatic feedback \cite{2011_Phys.Lett.A.375.788_Miller_et_al}. 

In this paper, we present a full non-linear treatment of this problem for the large-scale gravitational wave interferometers. A similar approach has been used by Polyakov and Vyatchanin in \cite{2007_Phys.Lett.A.368.423_Polyakov} to study the precursors of PI in the regime of input powers close to PI threshold. In this paper, we expand and generalise their treatment to arbitrary input power levels and investigate the time evolution of the amplitudes of Stokes and pump optical modes as well as of acoustic mode. \SD{Noteworthy is the fact that the set of dynamical equations in this work is similar to those in \cite{2007_Phys.Lett.A.368.423_Polyakov}, yet we prove that the requirement on smallness of the mechanical mode amplitude (allegedly, it has to be smaller than the optical modes linewidth) stated therein is not necessary, and this model is valid for arbitrarily large acoustic amplitudes.}

Besides, similar effect has been observed and analysed in small-scale whispering-gallery-mode optical resonators used to create stable radio-frequency optomechanical oscillators \cite{PhysRevLett.103.257403,Matsko:12}. Recently, the saturation of unstable oscillations has been observed at UWA and in LKB in tabletop optomechanical experiment using a high-finesse Fabry-Perot cavity with a silicon nitride membrane acting as the acoustic resonator test mass \cite{2014_arXiv1411.3016_Sundae_paper}. In that paper, we derived a similar model to back the experimental results. In this paper, we focus on a different physical system that comprises massive freely suspended Fabry-P\'erot cavities which interact with high-power optical field circulating inside. We also analyse the 3-mode parametric instability phenomenon in more detail and with greater generality.

The physics of three mode parametric instability is illustrated in Fig.~\ref{fig:PI_dynamics}. We consider a Fabry-Perot interferometer pumped at a laser frequency $\omega_p$ close to the resonance frequency, $\omega_0$, of one of the fundamental modes \cite{Kogelnik:66}. Ultrasonic vibrations of the mirrors with frequency $\Omega_m$ cause intracavity light to scatter into two motional sidebands, Stokes, with frequency $\omega_S = \omega_0-\Omega_m$, and anti-Stokes one, with frequency $\omega_{AS} = \omega_0+\Omega_m$. As the fundamental mode linewidth $\kappa_0$ is normally much smaller than $\Omega_m$, the amplitude of motional sidebands is normally not enhanced by the cavity. Not so, however, if the sideband frequency coincides with one of the higher order optical mode (HOM) frequencies. Such modes exist in any Fabry-Perot cavity and populate densely the free spectral range between the fundamental modes (see Sec. 3.3 of \cite{Kogelnik:66}). If the transverse spatial profile of such HOM matches the profile of an acoustic vibration, the photons scattered into this mode from the fundamental mode build up, thereby channeling part of the optical power circulating in the fundamental mode to the HOM. 

The beat note of the fundamental mode and the HOM, creates a near-resonant radiation pressure force on the mirror at the frequency of acoustic mode, $\Omega_m$. Now, there are two possibilities to consider. If HOM frequency coincides (approximately) with anti-Stokes sideband frequency, $\omega_{AS}$, the radiation pressure force will be applied out of phase with the acoustic vibrations, thereby damping them \cite{2002_PLA.299.326_Kells}. This effect is analogous to radiation pressure cooling \cite{PhysRevLett.92.075507,NJP.10.9.095007,PhysRevLett.97.243905,Nature.463.72,Nat.Phys.5.7.509_2009}, where the scattered Stokes photon energy is a sum of the fundamental mode energy and the acoustic mode phonon energy, \textit{i.e.} $\hbar\omega_{AS} = \hbar\omega_0+\hbar\Omega_m$. 

The instability we are studying in this paper, on the contrary, occurs when the HOM frequency matches the Stokes sideband frequency $\omega_S$: $\hbar\omega_{S} = \hbar\omega_0-\hbar\Omega_m$. To distinguish this HOM from others we will call it hereinafter a Stokes mode. In this case, radiation pressure force of the beat note is in phase with the acoustic oscillations, leading to amplification. This, in turn, makes amplitude of the Stokes sideband larger, thereby increasing the amplitude of the radiation pressure force. So, the loop closes and the instability breaks out. An illuminating description of this process in terms of feedback control theory can be found in the work of Evans \textit{et al.} \cite{2010_PLA_Evans_et_al}. 

The picture above gives no account for natural decay of the Stokes mode and acoustic oscillations due to loss, characterised by Stokes mode linewidth $\kappa_S$ Stokes and acoustic mode decay rate $\gamma_m$. These loss mechanisms counterbalance the instability, and for circulating power below the certain threshold (see below), there is no instability. However, as soon as the power circulating in the fundamental mode, reaches this threshold value, the 3-mode instability sets on. Very similar process happens in optical parametric oscillators, where above certain threshold pump power, in the presence of strong Kerr non-linearity of the medium, pump photons scatter into pairs of \textit{idler} and \textit{signal} photons \cite{1966_IEEE.J.Quant.El.2.9.418_Yariv_Louisell}.

This process resembles a relief valve operation, when above certain pressure the valve opens and redirects the excess fluid flow into a reserve pipe, keeping the main pipe pressure constant. Similarly, above the instability threshold, all the excess optical power is redirected from the fundamental mode to the Stokes mode and the acoustic mode oscillations. New steady state amplitudes of all three participating modes are reached when the balance of power is restored in the system, meaning that the amount of power pumped into the system matches the sum power leaving it through the three decay channels, characterised by the decay rates of the two optical modes, $\kappa_0$ and $\kappa_S$, and a mechanical decay rate, $\gamma_m$. 

In this paper, we present a non-linear theory of the 3-mode parametric instability in large scale gravitational wave interferometers with free suspended mirrors. We derive and solve equations of motion for 3 participating modes amplitudes and obtain their temporal dynamics. To characterise the behaviour of unstable modes, we calculate values of new steady-state amplitudes and give an estimate of instability development timescale. 

%This paper is organised as follows. In Sec.~\ref{sec:Model} we outline a theoretical model of the 3-mode optomechanical interaction. In Sec.~\ref{sec:TempDynamics} we solve for temporal dynamics of the 3 participating modes and derive ...

%\section{Previous studies}
%Here we shall give a brief review of all PI studies we know. 

\section{Model}\label{sec:Model}

\paragraph{Hamiltonian of 3-mode interaction.}
To represent parametric instability in large-scale GW interferometers, we use a simple model of a Fabry-P\'erot cavity with resonance frequency $\omega_0$ pumped with a laser, having frequency $\omega_p$ and power $P_{in}$. Below, we choose to adhere to Hamiltonian description of the system under study in contrast with the original work of Braginsky \textit{et al.} \cite{2001_Braginsky_PI_paper} where a completely equivalent Lagrangian approach has been used to derive equations of motion of interacting modes. It is worth emphasising that our analysis below is purely classical and no quantum effects are taken into account in this manuscript. Nevertheless, the choice of Hamiltonian classical description allows an easy expansion of this model to a quantum one. As a matter of fact, it has been done recently in the work that predicted a new source of quantum radiation pressure noise in gravitational-wave interferometers originating from the 3-mode optomechanical interaction \cite{2014_CQG.31.14.145002_Ju_Li}.

If we assume intracavity light to be linearly polarised and to propagate along the cavity optical axis $z$, its electric and magnetic field strain components can be represented in terms of expansion over cavity modes as:
\begin{subequations}
\begin{multline}
E(t,\,\vec{r}_\perp,\,z) = 
 \sum\limits_{J}\sqrt{\frac{\hbar\omega_J}{\epsilon_0 V}} f_J(\vec{r}_\perp)\sin(\omega_J z/c)\times\\
 \left[a_J(t)+a_J^\dag(t)\right]\,,
\end{multline}
\begin{multline}
H(t,\,\vec{r}_\perp,\,z) = 
  -i\sum\limits_{J}\sqrt{\frac{\hbar\omega_J}{\mu_0 V}} f_J(\vec{r}_\perp)\cos(\omega_J z/c)\times\\\left[a_J(t)-a_J^\dag(t)\right]\,,
\end{multline}
\end{subequations}
where $\hbar$, $c$, $\epsilon_0$ and $\mu_0$ stand for Plank constant, speed of light and vacuum permittivity and permeability, respectively; $V = L\mathcal{A}$ is the volume of the cavity of length $L$ occupied by a light beam with cross-section area $\mathcal{A}$; $f_J(\vec{r}_\perp)$ is  the J-th mode spatial distributions in the direction perpendicular to the propagation direction, $a_J$ denotes dimensionless complex amplitudes of the mode with frequency $\omega_J$ normalised so as $n_J = a_J^\dag a_J = |a_J|^2$ represents the number of photons in the corresponding mode. The acoustic mode of the mirror can be described in terms of surface deflection component along the z-axis:
$\zeta(\vec{r}_\perp,\,t) = x_{\rm z.p} u_z(\vec{r}_\perp)[b_m(t)+b_m^\dag(t)] $ with $u_z(\vec{r}_\perp)$ being the transverse mode spatial shape and $x_{\rm z.p.} = \sqrt{\hbar/(2 m \Omega_m)}$ is the ground state oscillations amplitude for an oscillator with effective mass $m$ and eigenfrequency $\Omega_m$.

Since parametric instability occurs only for the modes which satisfy a certain matching condition, $\omega_0 = \omega_S+\Omega_m+\Delta_m$ (detuning should be smaller than the larger of the Stokes or fundamental mode line width, $\Delta_m\ll\max[\kappa_S,\kappa_0]$), we can limit our consideration to those three modes (with $J=0,S$). The Hamiltonian for this 3-mode interacting system reads:
\begin{multline}
\mathcal{H} = \mathcal{H}_m- \\\frac{1}{2}\int\limits_{\mathcal{A}}d\vec{r}_\perp(L+\zeta)\left[\epsilon_0(E_0+E_S)^2+\mu_0(H_0+H_S)^2\right]\,,
\end{multline}
with $\mathcal{H}_m = \hbar\Omega_mb_m^\dag b_m$ and the last term describing the well known optomechanical interaction when radiation pressure force ($\propto$ light intensety) acts on a mechanical degree of freedom. After integration over transverse coordinates, $\vec{r}_\perp$, one can rewrite the it in a more familiar representation:
\begin{equation}\label{eq:Hamiltonian}
\mathcal{H} = \mathcal{H}_0 + \mathcal{H}_S + \mathcal{H}_m + \mathcal{H}_{\rm 0S}+ \mathcal{H}_{\rm 00}+ \mathcal{H}_{\rm SS} + \mathcal{H}_{\rm drive}
\end{equation}
where free evolution Hamiltonians for optical modes can be written as $\mathcal{H}_J = \hbar\omega_J a_J^\dag a_J$ ($J=0,S$), $\mathcal{H}_m = \hbar\Omega_m b_m^\dag b_m$, optomechanical interaction terms read
\begin{subequations}
\begin{eqnarray}
 \mathcal{H}_{\rm 0S} = -\hbar G_{0S}(b_m+b_m^\dag)(a_0+a_0^\dag)(a_S+a_S^\dag)\,, \\
 \mathcal{H}_{\rm 00} = -\hbar G_{00}(b_m+b_m^\dag)a_0^\dag a_0\,,\\
 \mathcal{H}_{\rm SS} = -\hbar G_{SS}(b_m+b_m^\dag)a_S^\dag a_S\,.
\end{eqnarray}
\end{subequations}
with optomechanical coupling strengths defined as
\begin{equation}\label{eq:G_IJ_def}
G_{IJ} = x_{\rm z.p.}\sqrt{\Lambda_{IJ}\omega_I\omega_J}/L\,,
\end{equation}
where $\Lambda_{IJ}$ is an overlap factor of spatial profiles of 3 participating modes, defined as
\begin{equation}\label{eq:Overlap_def}
\Lambda_{IJ} = [(L/V)\int\,d\vec{r}_\perp u_z(\vec{r}_\perp)f_I(\vec{r}_\perp)f_J(\vec{r}_\perp)]^2\,,
\end{equation}
 with $I,J = 0,S$.

To complete a picture we need to add a term responsible for coupling with the environment, $\mathcal{H}_{\rm ext}$ that can be expressed in terms of mode decay rates, $\kappa_{0,S}$ and $\gamma_m$, and corresponding external input fields, $\alpha^{in}_{0,S}$ and $\beta_{\rm th}$: $\mathcal{H}_{\rm ext} = \sum_{J=0,S}i\hbar\sqrt{\kappa_J}[a_J^\dag a^{in}_J+a_J(a_J^{in})^*] + i\hbar\sqrt{\gamma_m}[b_m^\dag b_{\rm th}+b_m b_{\rm th}^*]$. Here $a_0^{in} = (A_p+\delta a_0^{in})e^{-i\omega_pt}$ includes external laser pumping amplitude $A_p = \sqrt{P_{in}/(\hbar\omega_p)}$ and zero-mean fluctuations $a_0^{in}$, while two other modes are driven by fluctuations only. We assume optical mode fluctuations to be in vacuum state so as $\langle a^{in}_{0,S}(t)(a^{in}_{0,S}(t'))^\dag\rangle = \delta(t-t')$, and the mechanical damping noise, $b_{\rm th}$, corresponds to thermal white noise with correlation function $\langle b_{\rm th}(t)b^\dag_{\rm th}(t')\rangle = 2\gamma_mN_{\rm th}\delta(t-t')$, where $N_{\rm th} = (e^{\hbar\Omega_m/(k_BT)}-1)^{-1}$ is the average number of thermal phonons in the mechanical mode with $k_B$ Boltzmann's constant and $T$ mode temperature (usually, room temperature, $T=300$~K, is assumed). 

\paragraph{Natural scales and variable renormalisation.}
In this study, we are interested in classical large-amplitude dynamics of the system described by the above Hamiltonian which means high occupation numbers for all participating modes. Therefore, we change from quantum units to more physically sensible ones within the frames of this problem. For optical modes, a steady state amplitude of light in the fixed length Fabry-P\'erot cavity, $A_c = 2A_p/\sqrt{\kappa_0} = \sqrt{4P_{in}/(\hbar\omega_0\kappa_0)}\equiv \sqrt{\bar n_c}$, looks a natural scale. Mechanical mode can be scaled by a displacement amplitude $b_0$  necessary to shift the Stokes mode frequency by one half-linewidth, \textit{i.e.} $G_{0S} b_0 = \kappa_S/2$. The latter is a natural non-linearity scale of an optomechanical system that sets the applicability limit for linearised model thereof. 

Hereinafter we will operate with scaled amplitudes defined as:
\begin{align}\label{eq:normalisation}
& \alpha_{0,S} = a_{0,S}/A_c\,, & \beta_m = b_m/b_0 = 2G_{0S} b_m/\kappa_S\,.
\end{align}
Similar transformation has to be applied to noise terms.

\section{Temporal dynamics of parametric instability.}\label{sec:TempDynamics}
\paragraph{Equations of motion.}
One can now write down Heisenberg-Langevin equations for the system described by Hamiltonian \eqref{eq:Hamiltonian} in the frame rotating with each mode frequency. This means doing substitutions $a_{0,S}(t)\to a_{0,S}(t)e^{-i\omega_{0,S}t}$ and $b_m(t)\to b_m(t)e^{-i(\Omega_m+\Delta_m) t}$ (here we define frequency mismatch as $\Delta_m = \omega_0-\omega_S-\Omega_m$), and then dropping all the terms oscillating with frequency $\Omega_m$ and faster, one gets: 
\begin{subequations}\label{eq:PIeqns}
\begin{align}
\dot a_0 &= -\frac{\kappa_0}{2}a_0 +iG_{0S} a_S b_m + \sqrt{\kappa_0}(A_p+a_0^{in})\,,\label{eq:PIeqns_a}\\
\dot a_S &= -\left(\frac{\kappa_S}{2}+i\Delta_m\right) a_S +iG_{0S}a_0 b_m^\dag + \sqrt{\kappa_S}a_S^{in}\,,\label{eq:PIeqns_b}\\
\dot b_m &= -\frac{\gamma_m}{2}b_m +iG_{0S} a_0 a_S^\dag + \sqrt{\gamma_m}b_{\rm th}\,.\label{eq:PIeqns_c}
\end{align}
\end{subequations}
These equations can be rewritten in terms of dimensionless scaled amplitudes introduced in \eqref{eq:normalisation}:
\begin{subequations}\label{eq:PIeqns}
\begin{align}
\dot \alpha_0 &= -\frac{\kappa_0}{2}\alpha_0 +i\frac{\kappa_S}{2} \alpha_S \beta_m + \frac{\kappa_0}{2}+\sqrt{\kappa_0}\alpha_0^{in}\,,\label{eq:PIeqns_a}\\
\dot \alpha_S &= -\left(\frac{\kappa_S}{2}+i\Delta_m\right) \alpha_S +i\frac{\kappa_S}{2}\alpha_0 \beta_m^\dag + \sqrt{\kappa_S}\alpha_S^{in}\,,\label{eq:PIeqns_b}\\
\dot \beta_m &= -\frac{\gamma_m}{2}\beta_m +i\frac{\gamma_m}{2} \mathcal{R}_0\alpha_0 \alpha_S^\dag + \sqrt{\gamma_m}\beta_{\rm th}\,.\label{eq:PIeqns_c}
\end{align}
\end{subequations}
Here we introduced an important quantity, a parametric instability (PI) gain, $\mathcal{R}_0 = 4 G_{0S}^2\bar n_c/(\gamma_m\kappa_S)$ significance of which will be revealed below. Note that amplitude $\beta_m$ introduced by \ref{eq:normalisation} practically coincides with dimensionless mechanical amplitude $Z$ of \cite{2007_Phys.Lett.A.368.423_Polyakov}.

%It is reasonable to normalise the above optical and mechanical amplitudes with the appropriate scales. For optical modes the reasonable scale is the amplitude of the pumped fundamental mode before the onset of the instability, \textit{i.e.} $a_0(t)\to A_0\tilde{a}_0(t)$ and $a_S(t)\to A_0\tilde{a}_S(t)$ where $A_0 = 2 A_p/\sqrt{\kappa_0}$. Natural scale for the mechanical mode is the width of the cavity resonance in units of length, \textit{i.e.} $x_{\rm z.p.}b_m(t)\to (\lambda_p/\mathcal{F})\tilde{b}_m$ with $\lambda_0=2\pi c/\omega_p$ is carrier light wavelength and $\mathcal{F} = \pi c/(L\kappa_0)$ is cavity finesse.

As we are mostly interested in the strong signal dynamics of the 3-mode system, the noise terms in the above equations may be safely omitted. However, small initial nonzero amplitude of mechanical oscillations is necessary for nontrivial solution. Brownian thermal vibrations of the mirror provide this initial amplitude of $b_m$ which is equal to $\sqrt{N_{\rm th}} \simeq (k_B T/(\hbar\Omega_m))^{1/2}$.  Solving this system of equations numerically gives the characteristic result shown in Fig.~\ref{fig:PI_dynamics}.

%This result, albeit obtained for a single Fabry-P{\'e}rot cavity, depends only on 3 dimensionless quantities, namely the parametric gain $\mathcal{R}_0$, normalised detuning $\delta_m$ and optical linewidths ratio, $\kappa_0/\kappa_S$. As demonstrated in works \cite{Strigin200710, 2007_Quant.Electronics.37.1097_Strigin_Vyatchanin,2007_Phys.Lett.A.362.91_Gurkovsky,2009_CQG_Gras_et_al,2010_PLA_Evans_et_al} and as we show below in Sec.~\ref{sec:aLIGO_PI}, parametric instability in all, even so complicated optical setups as those of dual recycled second generation gravitational wave detectors, can be fully described by these three parameters. Therefore, our result is quite universal.

\begin{figure*}[ht]
 \includegraphics[width=.9\textwidth]{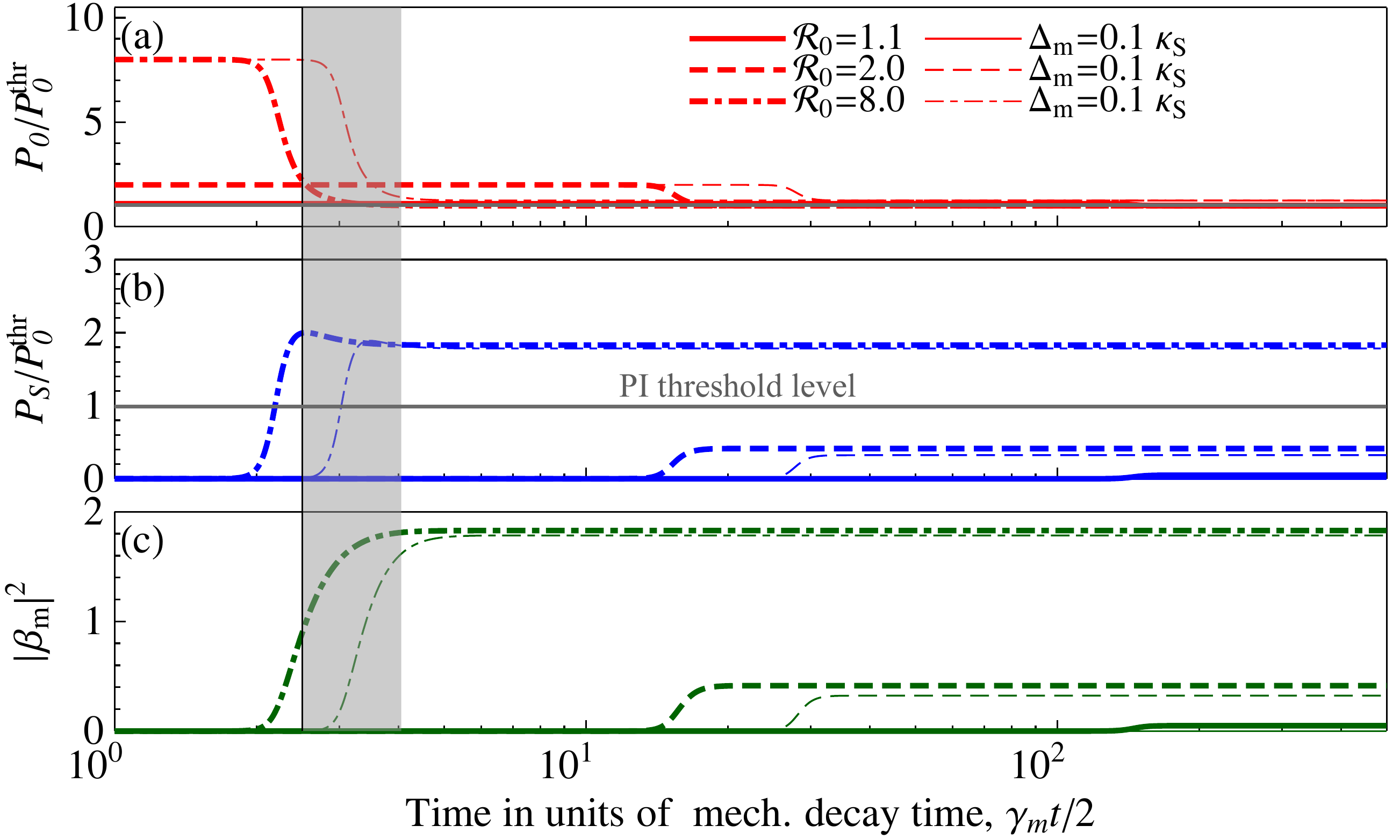}
 \caption{
 Temporal behaviour of 3 interacting modes for different values of PI gain $\mathcal{R}_0$. Thin lines show temporal dynamics in detuned case with frequency mismatch $\Delta_m = 0.1\kappa_S$. Optical modes amplitudes are normalised by maximal value of optical power circulating in fundamental mode which corresponds to PI gain value of $\mathcal{R}_0=8$. Acoustic mode amplitude is scaled by a displacement amplitude $b_0$ necessary to shift the Stokes mode frequency by one half-linewidth, as per definition in Eq.~\eqref{eq:normalisation}. Greyed out area and thin vertical line (indicating the maximum of the Stokes mode curve for $\mathcal{R}_0=8$) marks the time interval where inverse scattering process, $\hbar\omega_S+\hbar \Omega_m \to \hbar\omega_0$, intensifies in the system with PI gain larger than that defined by condition \eqref{eq:HOM_hump_cond}.}\label{fig:PI_dynamics}
 \end{figure*}

\paragraph{Parametric instability criterion.}
For parametric instability to develop, the certain threshold of input pumping power has to be reached. Braginsky \textit{et al.} \cite{2001_Braginsky_PI_paper} used the above introduced PI gain, $\mathcal{R}_0$ as a figure of merit for PI in the resonance case of $\Delta_m=0$ and defined it as:
\begin{align}\label{eq:PIcrit_res}
\mathcal{R}_0 &= \frac{4 G_{0S}^2\bar n_c}{\gamma_m\kappa_S} = \frac{8\Lambda_{0S}\omega_S P_{in}}{m\Omega_mL^2\gamma_m\kappa_S\kappa_0}\equiv \dfrac{P_{in}}{P_{in}^{\rm thres}} \,,\\
\mbox{where}\ & P_{in}^{\rm thres} = \frac{16 G^2_{0S}}{\hbar\omega_0\kappa_0\kappa_S\gamma_m} = \frac{m\Omega_mL^2\gamma_m\kappa_S\kappa_0}{8\Lambda_{0S}\nonumber\omega_S}
\end{align}
is the threshold input power value for resonant pumping.

If $\mathcal{R}_0\geqslant1$, the system goes unstable, otherwise no excitation of mechanical and Stokes mode occurs. In a more general, detuned case this criterion is only slightly modified:
\begin{equation}\label{eq:PIcrit_det}
\mathcal{R}_0\geqslant 1+\left(\dfrac{2\Delta_m}{\gamma_m+\kappa_S}\right)^2\,.
\end{equation}
This condition as well as an instability onset time can be obtained using simple linearised model based on Eqs.~\eqref{eq:PIeqns}. To start with we assume amplitude of a fundamental optical mode to be constant and equal to $1$ in the normalisation we chose.
%the steady state amplitude of a Fabry-P\'erot intacavity mode,  $A_0 = 2 A_p/\sqrt{\kappa_0}$. 
Then from Eqs.~\eqref{eq:PIeqns} we obtain the following set of linear equations for the Stokes and the mechanical modes:
\begin{subequations}
\begin{align}
\dot \alpha^\dag_S &= -\left(\frac{\kappa_S}{2}-i\Delta_m\right)\alpha^\dag_S -i\frac{\kappa_S}{2}\beta_m\,,\\
\dot \beta_m &= -\frac{\gamma_m}{2}\beta_m +i\frac{\gamma_m}{2}\mathcal{R}_0 \alpha_S^\dag\,.
\end{align}
\end{subequations}
%\begin{subequations}
%\begin{align}
%\dot a^\dag_S &= -\frac{\kappa_S}{2}a^\dag_S +iG_{0S}A_0^* b_m\,,\\
%\dot b_m &= -\left(\frac{\gamma_m}{2}+i\Delta_m\right)b_m +iG_{0S} A_0 a_S^\dag\,.
%\end{align}
%\end{subequations}
Looking for general solution in the form $\{\alpha_S^\dag,\,\beta_m\}\propto e^{(\Gamma+i\nu)t}$, the PI condition, $\Gamma\geqslant0$, is obtainable from the characteristic equation for the above linear system:
$$(\Gamma+i\nu+\kappa_S/2-i\Delta_m)(\Gamma+i\nu+\gamma_m/2)-\gamma_m\kappa_S\mathcal{R}_0/4=0$$
with the solution:
\begin{subequations}
\begin{align}
\Gamma &= \frac14\left[\sqrt{X+\sqrt{X^2+Y^2}}-(\kappa_S+\gamma_m)\right]\,,\label{eq:PIGamma}\\ 
\nu &= \frac{\Delta_m}{2}-\frac{1}{4}\sqrt{\sqrt{X^2+Y^2}-X}\,,\label{eq:PInu}
\end{align}
\end{subequations}
where
$$X\equiv 2\gamma_m\kappa_S\mathcal{R}_0-2\Delta_m^2+\frac12(\kappa_S-\gamma_m)^2\,,\quad Y\equiv2\Delta_m(\kappa_S-\gamma_m)\,.$$
Requirement $\Gamma>0$ yields the sought for relations \eqref{eq:PIcrit_res} and \eqref{eq:PIcrit_det} .

\section{Adiabatic elimination of cavity modes.}
In gravitational-wave interferometers, as well as in small-scale, table-top optomechanical experiments \cite{2014_arXiv1411.3016_Sundae_paper}, mechanical decay rate, $\gamma_m$, is way smaller than the decay rates of the optical degrees of freedom, $\kappa_{0,S}\gg\gamma_m$. Therefore one can safely assume optical modes to follow any changes in the mechanical mode almost instantaneously, without delay. Hence we can adiabatically eliminate optical modes by setting time derivatives in Eqs.~\eqref{eq:PIeqns_a} and \eqref{eq:PIeqns_b} to zero and express the two optical modes as functions of mechanical amplitude $\beta_m(t)$ as:
\begin{align}
\alpha_0(t) &= \frac{1+i\delta_m}{1+i\delta_m+\frac{\kappa_S}{\kappa_0}|\beta_m(t)|^2}\,, \label{eq:FMamp}\\ 
\alpha_S(t) &= \frac{i\beta_m^\dag(t)}{1+i\delta_m+\frac{\kappa_S}{\kappa_0}|\beta_m(t)|^2} \label{eq:HOMamp}\,,
\end{align}
where we defined dimensionless detuning $\delta_m=2\Delta_m/\kappa_S$. Substituting these expressions into the last Eq.~\eqref{eq:PIeqns_c}, one arrives at a single non-linear differential equation for the mechanical amplitude $\beta_m$:
\begin{equation}\label{eq:MampODE}
\dot \beta_m +\frac{\gamma_m}{2}\left(1-\frac{\mathcal{R}_0(1+i\delta_m)}{(1+\frac{\kappa_S}{\kappa_0}|\beta_m|^2)^2+\delta_m^2}\right)\beta_m =0\,.
\end{equation}
This equation displays the mechanism of saturation clearly. Indeed, the system gets unstable when the real part of the expression in brackets turns negative, meaning negative mechanical decay rate. However, the rise of amplitude, $\beta_m$, entering the denominator renders the negative term smaller and smaller, eventually reaching the critical point when the bracket turns 0. From this ensues a new, non-linear parametric instability condition of the form:
\begin{equation}
\mathcal{R}_{\rm NL} = \frac{\mathcal{R}_0}{\left(1+ \frac{\kappa_S}{\kappa_0}|\beta_m|^2\right)^2+\left(\frac{2\Delta_m}{\kappa_S}\right)^2}\geqslant 1\,,
\end{equation}
where the non-linear gain, $\mathcal{R}_{\rm NL}$, gradually wanes as amplitude, $\beta_m$, waxes, reaching a limit cycle. 

The value of the critical amplitude is identical to the steady state amplitude, as $\beta_m(t)$ is a monotone function of time and equals to:
\begin{equation}\label{eq:MampSS}
\bar\beta_m = \left[\frac{\kappa_0}{\kappa_S}(\sqrt{\mathcal{R}_0-\delta_m^2}-1)\right]^{1/2}\,.
\end{equation}
Steady state amplitudes for optical modes immediately ensue from Eqs~\eqref{eq:HOMamp} and \eqref{eq:FMamp}:
\begin{align}\label{eq:OMampsSS}
\bar\alpha_0 &= \sqrt{\frac{1+\delta_m^2}{\mathcal{R}_0}}\,, & \bar\alpha_S &= \left[\frac{\kappa_0}{\kappa_S}\frac{1}{\mathcal{R}_0} (\sqrt{\mathcal{R}_0-\delta_m^2}-1)\right]^{1/2}\,.
\end{align}

Adiabatic limit can also help to understand the small hump one can notice on the plot of the Stokes mode power in Fig.~\ref{fig:PI_dynamics} for PI gain $\mathcal{R}_0=8$ which is absent on the other curves with lower gain. The existence of this hump stems from the dependence of the Stokes mode amplitude, $\alpha_S$, on the mechanical amplitude, $\beta_m$, given by Eq.~\eqref{eq:HOMamp}. If we consider the Stokes mode normalised power $|\alpha_S|^2$ and equate its time derivative to zero we get:  
\begin{equation*}
 \partial_t(|\beta_m|^2)\cdot\dfrac{1+\delta_m^2-\frac{\kappa_S^2}{\kappa_0^2}|\beta_m|^2}{(1+\frac{\kappa_S}{\kappa_0} |\beta_m|^2)^2+\delta_m^2}=0\,,
\end{equation*}
which is a necessary condition for $|\alpha_S|^2$ to reach its extremum. As we know from numerical solution, $\beta_m(t)$ is monotonic, and $\partial_t(|\beta_m|^2)=0$ means mechanical amplitude has reached its maximal steady state value $\bar\beta_m$. Therefore, the condition for the hump would be that the second term in the product above becomes equal to zero for $\beta_m<\bar \beta_m$ that, accounting for the monotonic character of $\beta_m(t)$, means that the numerator of the second term is smaller than zero at $\bar \beta_m$.
From this immediately ensues a condition on parametric gain value above which one shall observe this hump in temporal behaviour of the Stokes mode:
\begin{equation}\label{eq:HOM_hump_cond}
\mathcal{R}_0 >\left[\frac{\kappa_0}{\kappa_S}(2+\delta_m^2)\right]^2+\delta_m^2\underset{\delta_m=0, \kappa_0=\kappa_S} {\longrightarrow}4 \,.
\end{equation}

Remarkably, in the resonance case, the value of PI gain $\mathcal{R}_0 = 4$, above which the hump develops in the Stokes mode corresponds to the situation when the power circulating in the Stokes mode gets equal to the threshold value, \textit{i.e.} to the power circulating in the fundamental optical mode. The hump represents the transient process of the initial excess growth of the Stokes mode occupation number above the threshold value before the acoustic mode could reach its steady state occupation number (recall that $\kappa_S\gg\gamma_m$) and further release of this level through inverse scattering to the carrier mode. The latter process invokes recombination of the excess $\omega_S$ photons with acoustic phonons at $\Omega_m$ yielding generation of $\omega_0$ photons. As shown by greyed out region in Fig.~\ref{fig:PI_dynamics}, when $\mathcal{R}_0=8$, the decrease of the fundamental mode power and the growth of the acoustic mode amplitude near the time when the Stokes mode maximum is reached gets slower which is indicative of the intensification of inverse scattering.

\begin{figure}[ht]
 \includegraphics[width=.49\textwidth]{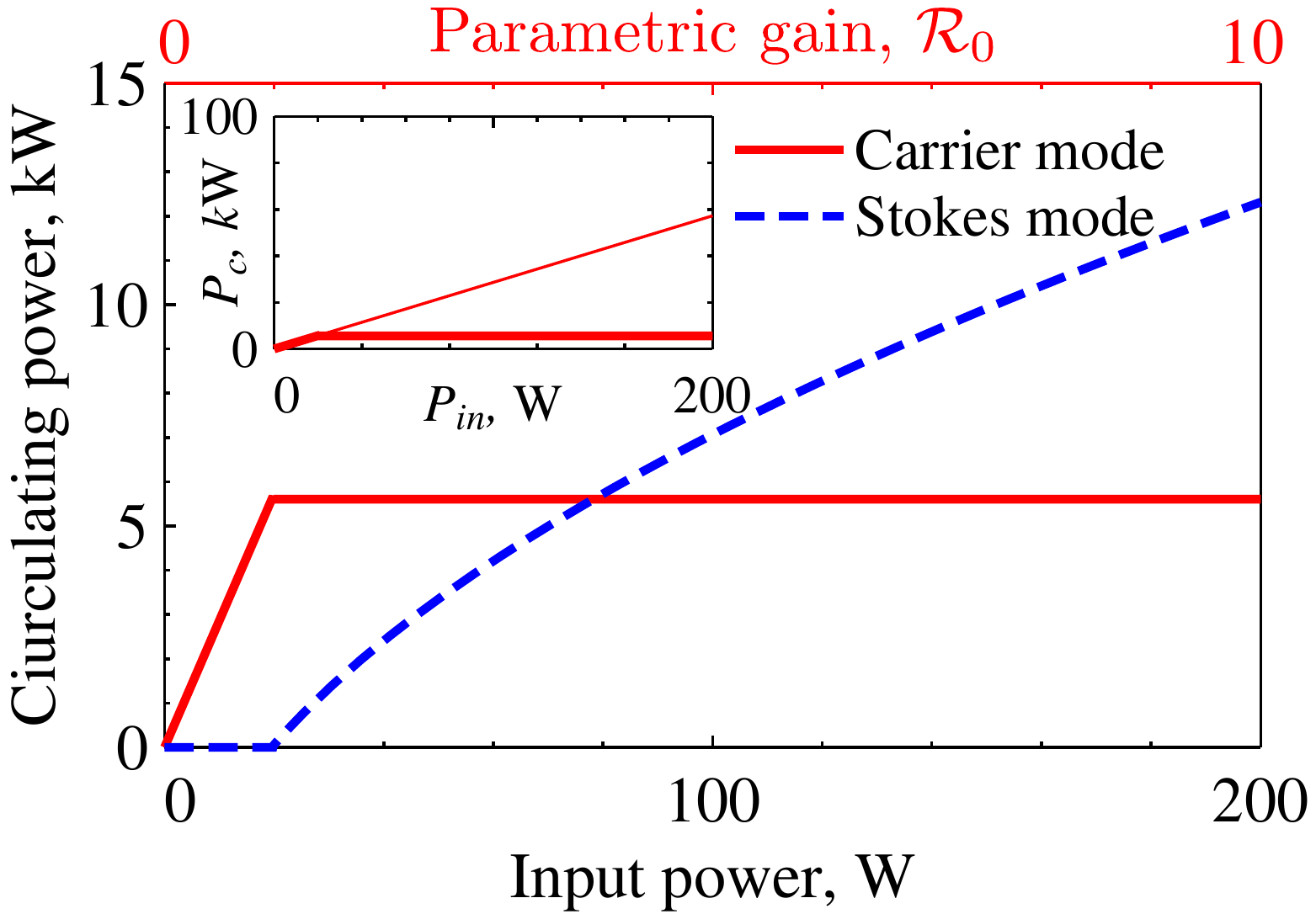}\\
 \includegraphics[width=.49\textwidth]{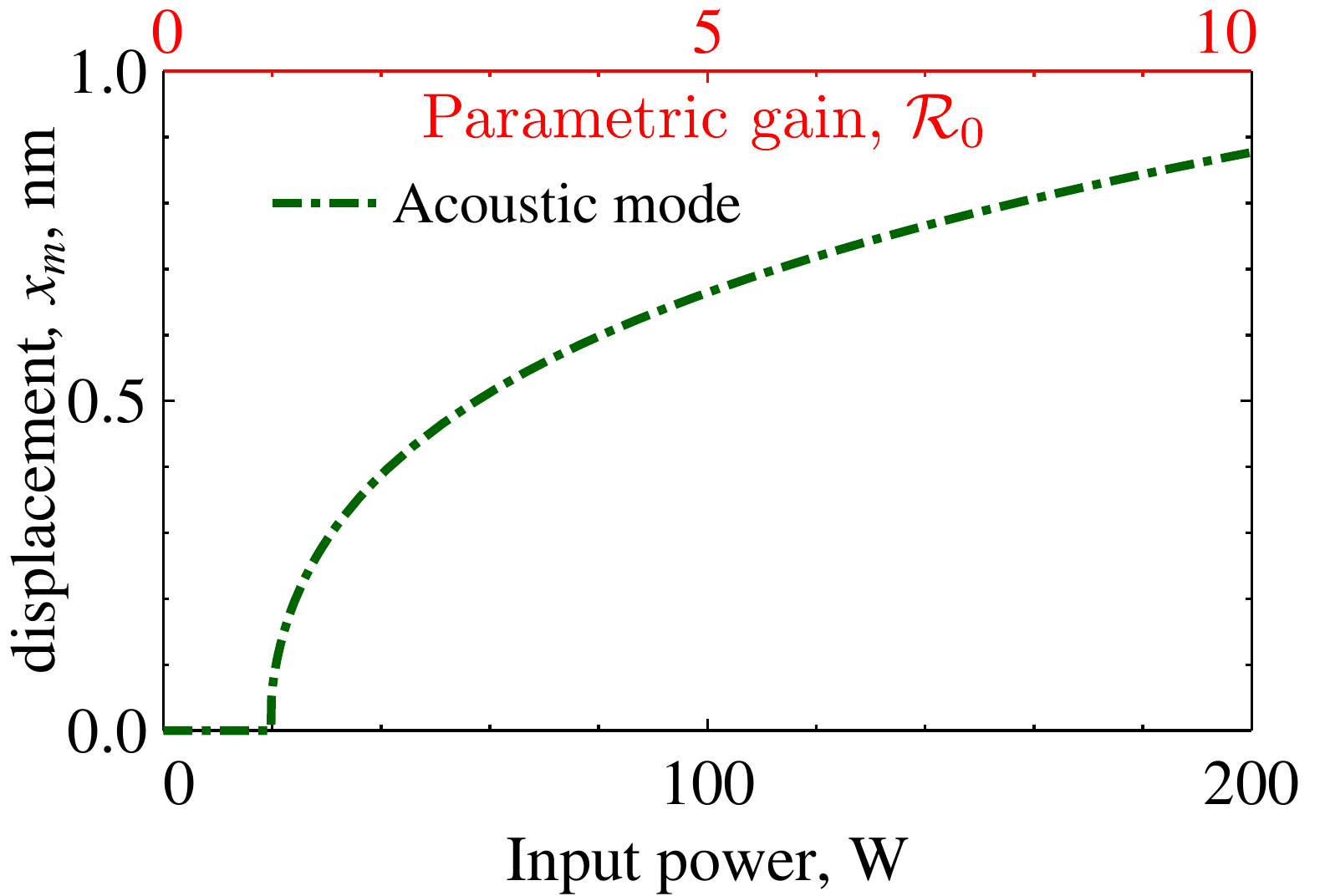}
 \caption{
  \textit{Upper panel: } Dependence of power circulating in fundamental mode (red solid line) and in higher-order Stokes mode (blue dashed line) on input power, calculated from expressions (\ref{eq:P_0_SS}--\ref{eq:P_m_SS}), using parameters from Table~\ref{tab:params}. Inset plot compares the behaviour of circulating power with (thick red line) and without (thin red line) PI.
  \textit{Lower panel:} Dependence of acoustic mode amplitude on input laser power for parameters listed in Table~\ref{tab:params}.}\label{fig:PI_SS_amps}
\end{figure}

%of light power circulating in optical modes and of displacement amplitude in metres:
%%\begin{subequations}
%\begin{align}\label{eq:SS_physical}
%x_m^2 &= \frac{L^2\kappa_0\kappa_S}{4\Lambda_{0S}\omega_0\omega_S}\left[\sqrt{\frac{8\Lambda_{0S}\omega_S P_{in}}{m\Omega_m L^2\gamma_m\kappa_S\kappa_0} -\frac{4\Delta_m}{\kappa_S^2}}-1\right],\nonumber\\
%P_0 &=  \frac{m\Omega_m c L\gamma_m\kappa_S}{4\Lambda_{0S}\omega_S}\left[1+\frac{4\Delta_m^2}{\kappa_S^2}\right]\,,\\
%P_S &= \frac{m\Omega_m c L\gamma_m\kappa_0}{4\Lambda_{0S}\omega_S}\left[\sqrt{\frac{8\Lambda_{0S}\omega_S P_{in}}{m\Omega_m L^2\gamma_m\kappa_S\kappa_0} -\frac{4\Delta_m}{\kappa_S^2}}-1\right]\nonumber \,.
%\end{align}
%%\end{subequations}
%Taking into account that threshold power circulating in the fundamental mode is equal to
%\begin{equation}\label{eq:P_thres_cirrc}
%P_0^{thres} = \frac{2 c}{\kappa_0 L}P_{in}^{thres} = \frac{m\Omega_m c L\gamma_m\kappa_S}{4\Lambda_{0S}\omega_S}\,,
%\end{equation}
%the above expressions can be rewritten in a more compact form:
%\begin{align}\label{eq:SS_physical}
%x_m^2 &= \frac{L^2\kappa_0\kappa_S}{4\Lambda_{0S}\omega_0\omega_S}\left[\sqrt{\frac{P_c}{P_0^{thres}} -\frac{4\Delta_m}{\kappa_S^2}}-1\right],\nonumber\\
%P_0 & = P_0^{thres}\left[1+\frac{4\Delta_m^2}{\kappa_S^2}\right]\,,\\
%P_S &= \frac{\kappa_0}{\kappa_S}P_0^{thres}\left[\sqrt{\frac{P_c}{P_0^{thres}} -\frac{4\Delta_m}{\kappa_S^2}}-1\right]\nonumber \,.
%\end{align}

\paragraph{Approximate solution.}
 By representing a complex function $\beta_m(t)$ as $|\beta_m(t)|e^{i\phi_m(t)}$ one can easily obtain from \eqref{eq:MampODE} the equations they satisfy:
\begin{subequations}
\begin{align}
|\dot\beta_m| &= -\frac{\gamma_m}{2}|\beta_m|\left(1-\frac{\mathcal{R}_0}{(1+\frac{\kappa_S}{\kappa_0}|\beta_m|^2)^2+\delta_m^2}\right) \,,\label{eq:MampmodDE}\\
\dot\phi_m &= -\frac{\gamma_m}{2}\frac{\delta_m\mathcal{R}_0}{(1+\frac{\kappa_S}{\kappa_0}|\beta_m|^2)^2+\delta_m^2}\,.\label{eq:MampArgDE}
\end{align}
\end{subequations}
One can solve this system of first-order differential equations explicitly, but the result will be an implicit transcendental equation on $\beta_m$ that cannot be resolved analytically. Nevertheless, a pretty good approximation thereto gives the solution of a simplified equation for $\beta_m$, obtainable by expanding the bracket in the RHS of \eqref{eq:MampmodDE} in $|\beta_m|$ around the $\bar\beta_m$ of \eqref{eq:MampSS}. This results in a Bernoulli equation \cite{2003PolyaninODEref} of the shape:
\begin{align*}
|\dot\beta_m| &= \mathcal{D}|\beta_m|\left(1-\frac{|\beta_m|}{\bar\beta_m}\right)\,, & \mathcal{D} &= \frac{2\gamma_m\kappa_S\bar\beta_m^2}{\kappa_0\mathcal{R}_0} \left(1+\frac{\kappa_S}{\kappa_0}\bar\beta_m^2\right)\,,  
\end{align*}
that, accounting for initial condition $|\beta_m(0)|$ resolves in:
\begin{equation}
|\beta_m(t)| = \frac{|\beta_m(0)|\bar\beta_me^{\mathcal{D}t}}{\bar\beta_m+|\beta_m(0)|(e^{\mathcal{D}t}-1)}\,.
\end{equation}

\begin{table}[h]
%\squeezetable
     \begin{ruledtabular}
     	\begin{tabular}{l	 c c}
     		Parameter & Notation & Value \\\hline
		Effective mass, kg & $m$ & 40 \\
		Arm length, km & $L$ & $4$\\
		Overlap factor & $\Lambda$ & $1.0$\\
		Fund. mode finesse & $\mathcal{F}_0$ & 450\\
		Stokes mode finesse & $\mathcal{F}_S$ & 450\\
		Mech. frequency, kHz & $\Omega_m/2\pi$ & $20$\\
		Mech. Q & $Q_m$ & $10^7$\\
		Temperature, K & $T$ & $300$\\\hline
		PI gain & $\mathcal{R}_0$ & $0.05 \left(\frac{P_{in}}{1\mathrm{\ W}}\right)$
	\end{tabular}
     \end{ruledtabular}
	\caption{Parameters used for simulation}
	\label{tab:params}
\end{table}

\section{Power relations in 3-mode system}

It is obvious that in adiabatic limit ($\Omega_m\ll\omega_{0,S}$), a sum of optical powers leaving the interferometer in the fundamental and the Stokes' modes, \textit{i.e.} $P^{out}_0+P^{out}_S$,  must be equal to the power $P_{in}$ entering it. Power scattered in the acoustic mode is negligible compared to that of the optical modes due to a huge frequency difference. In order to see that it is indeed the case for our treatment one needs to write down standard input-output relations connecting light amplitudes outside the interferometer (of incident and reflected beams) with the intracavity ones derived in \eqref{eq:FMamp} and\eqref{eq:HOMamp}:
\begin{align}
a^{in}_0 + a^{out}_0 &= \sqrt{\kappa_0} a_0 & \Rightarrow\ A_p+A^{out}_0 = \sqrt{\kappa_0}A_0 \,,\\
a^{in}_S + a^{out}_S &= \sqrt{\kappa_S} a_S & \Rightarrow\ A^{out}_S = \sqrt{\kappa_S}A_S \,,
\end{align}
where capital letters identify classical components of the light fields we are only concerned with in this work. Powers in the corresponding beams are related to these amplitudes as $P_J = \hbar\omega_J |A_J|^2$ ($J$ stands for $p,0,S$). 
Assuming steady-state, \textit{i.e.} setting $\beta_m(t)\to\bar\beta_m$  in \eqref{eq:FMamp} and\eqref{eq:HOMamp}, and  normalisation relations \eqref{eq:normalisation}, one gets expressions for the reflected light powers as follows:
\begin{align*}
P^{out}_0 &= P_{in} \dfrac{(1-\frac{\kappa_S}{\kappa_0}|\bar\beta_m|^2)^2+\delta_m^2}{(1+\frac{\kappa_S}{\kappa_0}|\bar\beta_m|^2)^2+\delta_m^2}\,,\\ 
P^{out}_S &= P_{in} \dfrac{4\frac{\kappa_S}{\kappa_0}|\bar\beta_m|^2}{(1+\frac{\kappa_S}{\kappa_0}|\bar\beta_m|^2)^2+\delta_m^2}\,.
\end{align*}
\SD{There is no problem to see now that $P^{out}_0+P^{out}_S = P_{in}$ and the power balance is observed for any value of the acoustic mode amplitude $\bar \beta_m$. Substituting Eq.~\eqref{eq:MampSS} in these equations, one gets the nonlinear I/O-relations for the 3-mode system above the PI threshold:
\begin{align}
P^{out}_0 &= P_{in} \left[1-4\dfrac{\sqrt{\mathcal{R}_0-\delta_m^2}-1}{\mathcal{R}_0}\right]\,,\\
P^{out}_0 &= 4 P_{in} \dfrac{\sqrt{\mathcal{R}_0-\delta_m^2}-1}{\mathcal{R}_0}\,,\ \mathcal{R}_0 = \frac{P_{in}}{P_{in}^{\rm thr}}\,.
\end{align}
The dependence of outgoing power for each mode on PI gain and thereby on input power is drawn in Fig.~\ref{fig:Pout_vs_Pin}.

\begin{figure}
\includegraphics[width=0.5\textwidth]{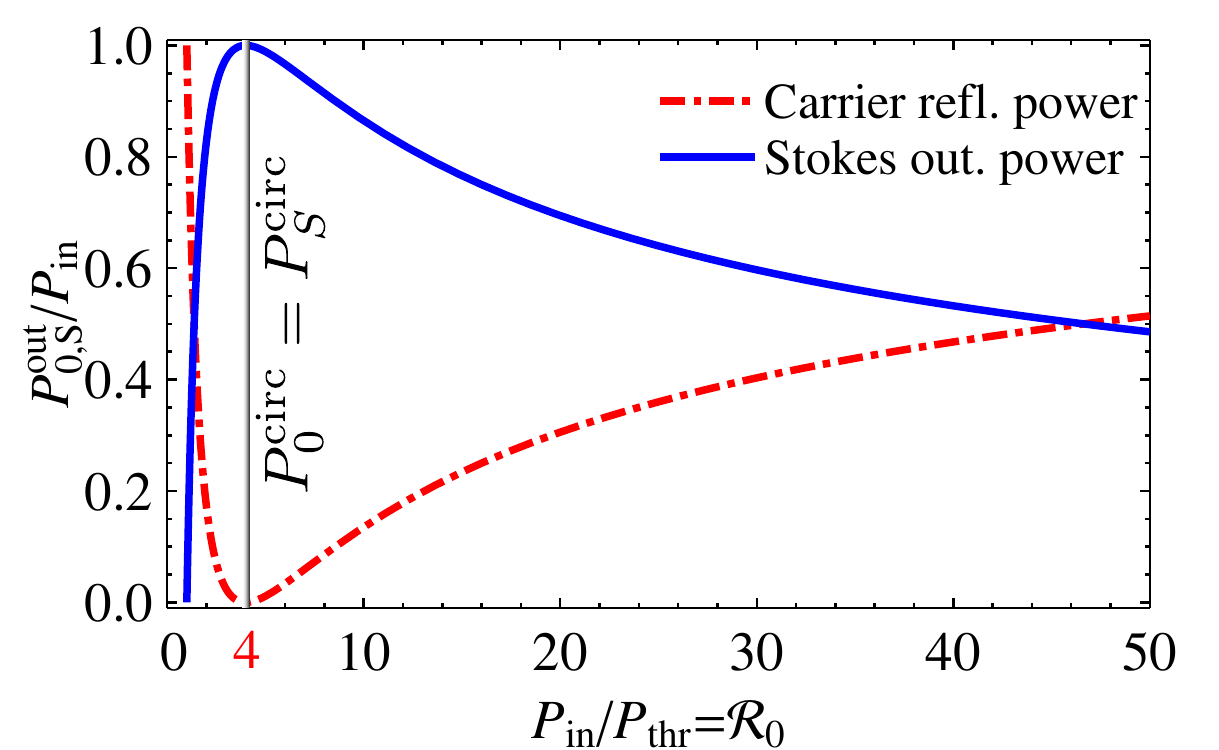}
\caption{\SD{Dependence of reflected light power at the carrier frequency $\omega_0$ (red dash-dotted line) and at the HOM frequency (blue line) as a function of parametric gain $\mathcal{R}_0$ and, thereby of input power. Note the non-linear and non-monotonic relation between input and reflected powers. Critical PI gain value of $\mathcal{R}_0=4$ corresponds to the case of critical coupling, \textit{i.e.} when the non-linear loss of carrier photons to the Stokes mode and to the acoustic mode reaches the level of bare loss of the interferometer defined by the reflectivities of the mirrors.}}\label{fig:Pout_vs_Pin}
\end{figure}
The interesting feature of these relation is their apparent non-linearity that is quite opposite to a naive assumption one might be tempted to make that intracavity circulating power in both optical modes is linearly proportional to the outgoing power. Note also the existence of critical value of input power corresponding to PI gain of $\mathcal{R}_0 = 4$ (in resonance case) when the reflected power in carrier mode vanishes and the Stokes mode output reaches its maximum. Physics of this process is straightforward, for it is this level of power when the loss of carrier photons due to 3-mode scattering (into Stokes and acoustic modes) reaches the level of the cavity bare loss, summarised in $\kappa_0$. This is a well known phenomenon of critical coupling. This, in particular, has profound observational consequences, as measuring the PI by recording the beat note of the reflected lights at the carrier and Stokes mode frequencies may result in zero signal for the range of PI gains around the critical one, \textit{i.e.} around $\mathcal{R}_0\simeq 4$.  }

%To some extent, one may say that the losses of the fundamental mode in the 3-mode interferometer depend on the incident power in a non-linear way. This is simply a manifestation of the fact that part of the intracavity power is scattered into the Stokes and acoustic modes by means of the 3-mode interaction process.  

Returning to intracavity fields and recalling the definition of normalised modes \eqref{eq:normalisation}, one can rewrite expressions \eqref{eq:MampSS} and \eqref{eq:OMampsSS} in more physical terms as: 
\begin{align}
\dfrac{P_0}{\omega_0} &= \dfrac{P^{\rm thr}_0}{\omega_0}\left[1+\frac{4\Delta_m^2}{\kappa_S^2}\right]\,,\label{eq:P_0_SS}\\
\dfrac{P_S}{\omega_S} &= \dfrac{P^{\rm thr}_0}{\omega_0}\left[\sqrt{\frac{P_{in}}{P_{in}^{\rm thr}} -\frac{4\Delta_m^2}{\kappa_S^2}}-1\right]\,,\label{eq:P_S_SS}\\
\dfrac{P_m}{\Omega_m} &= \dfrac{P^{\rm thr}_0}{\omega_0}\left[\sqrt{\frac{P_{in}}{P_{in}^{\rm thr}} -\frac{4\Delta_m^2}{\kappa_S^2}}-1\right]\,, \label{eq:P_m_SS}
\end{align}
where 
\begin{equation}\label{eq:PI_thres_circ}
P^{\rm thr}_0 = \hbar\omega_0\dfrac{\gamma_m\kappa_0\kappa_S}{4 G_{0S}^2} = \frac{m\Omega_m L^2\gamma_m\kappa_0\kappa_S}{2\Lambda_{0S}\omega_S}\,,
\end{equation}
is the power circulating in the fundamental mode at resonance frequency $\omega_0$ when PI threshold is reached. The plots of these expressions are shown in Fig.~\ref{fig:PI_SS_amps}. Here mechanical amplitude is inferred from the acoustic power using its relation to acoustic amplitude: $P_m = \gamma_m \overline{\mathcal{H}}_m = \gamma_m m\Omega_m^2 \overline{x_m^2}$, yielding:
\begin{equation}
\overline{x_m^2} = \frac{L^2\kappa_0\kappa_S}{2\Lambda_{0S}\omega_0\omega_S}\left[\sqrt{\frac{P_{in}}{P_{in}^{\rm thr}} -\frac{4\Delta_m^2}{\kappa_S^2}}-1\right]\,.
\end{equation}

The above equations \eqref{eq:P_S_SS} and \eqref{eq:P_m_SS} have another interesting implication, namely the equality
\begin{equation}\label{eq:Manley-Rowe}
\dfrac{P_S}{\omega_S} = \dfrac{P_m}{\Omega_m}
\end{equation}
represents the well known Manley-Rowe relations for a non-linear interacting systems \cite{2003bookBoyd,1956_Proc.IRE.44.904_Manley-Rowe,1958_Proc.IRE.46.850_Rowe}. In our case it says that the number of Stokes photons produced from the $\omega_0$ photons matches exactly the number of acoustic phonons generated in this process.  

One might wonder if there is an inverse process going on in the system, \textit{i.e.} the generation of $\omega_0$ photons from the pairs of $\omega_S$ photons and $\Omega_m$ phonons, and if this process shall prevent the power circulating in the Stokes mode to exceed that in the fundamental mode. Indeed, such inverse scattering may happen and, were it not for constant pumping of laser photons into the system at frequency $\omega_0$, there will be equal probability for direct and inverse scattering processes leading to equilibrium between the occupation numbers of the modes. One has to remember about losses that constantly drain photons and phonons from the corresponding modes. It is these losses that the PI threshold \eqref{eq:PIcrit_res} owes its existence to.    

The threshold of PI represents the level of intracavity power, whereat no more $\omega_0$ photons can be born by the fundamental mode. Let the power in the Stokes mode reach the level slightly higher than the threshold power as represented by the crossing of blue and red lines in the upper panel of Fig.~\ref{fig:PI_SS_amps}. It means that input power is more than 4 times higher than $P_{in}^{\rm thr}$. The inverse scattering process creates then a photon at $\omega_0$, thereby making fundamental mode to have one photon more than the threshold allows. This photon cannot decay away using the fundamental mode loss channel, as it is saturated at the 4 times lower input power level of $P_{in}^{\rm thr}$. The only way for it to escape is through scattering to the Stokes photon and acoustic phonon again. The probability of this scattering is higher than of escaping the cavity because the optoacoustic photon-phonon exchange rate $G_{0S}\bar n_c^{1/2}$ is faster than the cavity decay rate $\kappa_0$, which is the prerequisite for parametric instability to start in the first place.

\section{Timescale of instability: onset time.}\label{sec:PIonsetT}
Using expression \eqref{eq:PIGamma} for $\Gamma$, one can derive a timescale for the instability onset, which is a timescale whereat linearised model breaks down and exponential ring-up gives place to a saturation and, eventually, to a new equilibrium state reached by a system. It can be estimated as a moment when exponentially growing mechanical amplitude, $\beta_m(t)\simeq \beta_m(0)e^{\Gamma t}$, with $\beta_m(0) = \sqrt{4 G_{0S}^2 N_{th}/\kappa_S^2}$, reaches the above calculated steady state level, $\bar\beta_m$ of Eq.~\eqref{eq:MampSS}, \textit{i.e.}:
%One can define this moment as a time when the parametric frequency shift produced by the mechanical oscillations hits the Stokes mode linewidth, \textit{i.e.} :
\begin{multline}\label{eq:PI_onset_time_exact}
|\beta_m(0)|e^{\Gamma \tau_{\rm PI}} = \bar\beta_m\ \Rightarrow\\  
\tau_{\rm PI} = \frac{1}{2\Gamma}\log\dfrac{\kappa_0\kappa_S(\sqrt{\mathcal{R}_0-\delta_m^2}-1)}{4G^2_{0S} N_{\rm th}} =\\ \frac{1}{2\Gamma}\log\dfrac{m\Omega_m^2 L^2\kappa_0\kappa_S(\sqrt{\mathcal{R}_0-(2\Delta_m/\kappa_S)^2}-1)}{2\Lambda_{0S}\omega_0\omega_S k_B T}
%\frac{\kappa_S}{2 G_{0S}\sqrt{N_{\rm th}}} = \frac{1}{2\Gamma}\log\frac{\kappa_S\bar n_0}{\gamma_m N_{\rm th}}\,.  
\end{multline}
This expression can be simplified if we recall that in real interferometers $\gamma_m\ll\kappa_S$ and condition $\Delta_m\ll\kappa_S$ should be satisfied for PI to arise. Expanding Eq.~\eqref{eq:PIGamma} in Taylor series in $\gamma/\kappa$ and $\Delta_m/\kappa_S$, one gets:
\begin{equation*}
\Gamma\simeq \frac{\gamma_m \mathcal{R}_0}{2}[1-\mathcal{R}_0^{-1}-\delta_m^2]\,.
\end{equation*}
Thus one can get the following approximate expression for PI onset time:
\begin{multline}\label{eq:PI_onset_time_approx}
\tau_{\rm PI}\simeq \frac{1-\mathcal{R}_0^{-1}+\delta_m^2}{\gamma_m \mathcal{R}_0}\log\dfrac{\bar{n}_c\kappa_0}{\bar N_{th}\gamma_m}\dfrac{\sqrt{\mathcal{R}_0-\delta_m^2}-1)}{\mathcal{R}_0} = \\
= \frac{1-\mathcal{R}_0^{-1}+\delta_m^2}{\gamma_m \mathcal{R}_0}\log\dfrac{4P_{in}\Omega_m}{k_BT \omega_0\gamma_m}\dfrac{(\sqrt{\mathcal{R}_0-\delta_m^2}-1)}{\mathcal{R}_0}
\end{multline}

\begin{figure}[ht]
 \includegraphics[width=.49\textwidth]{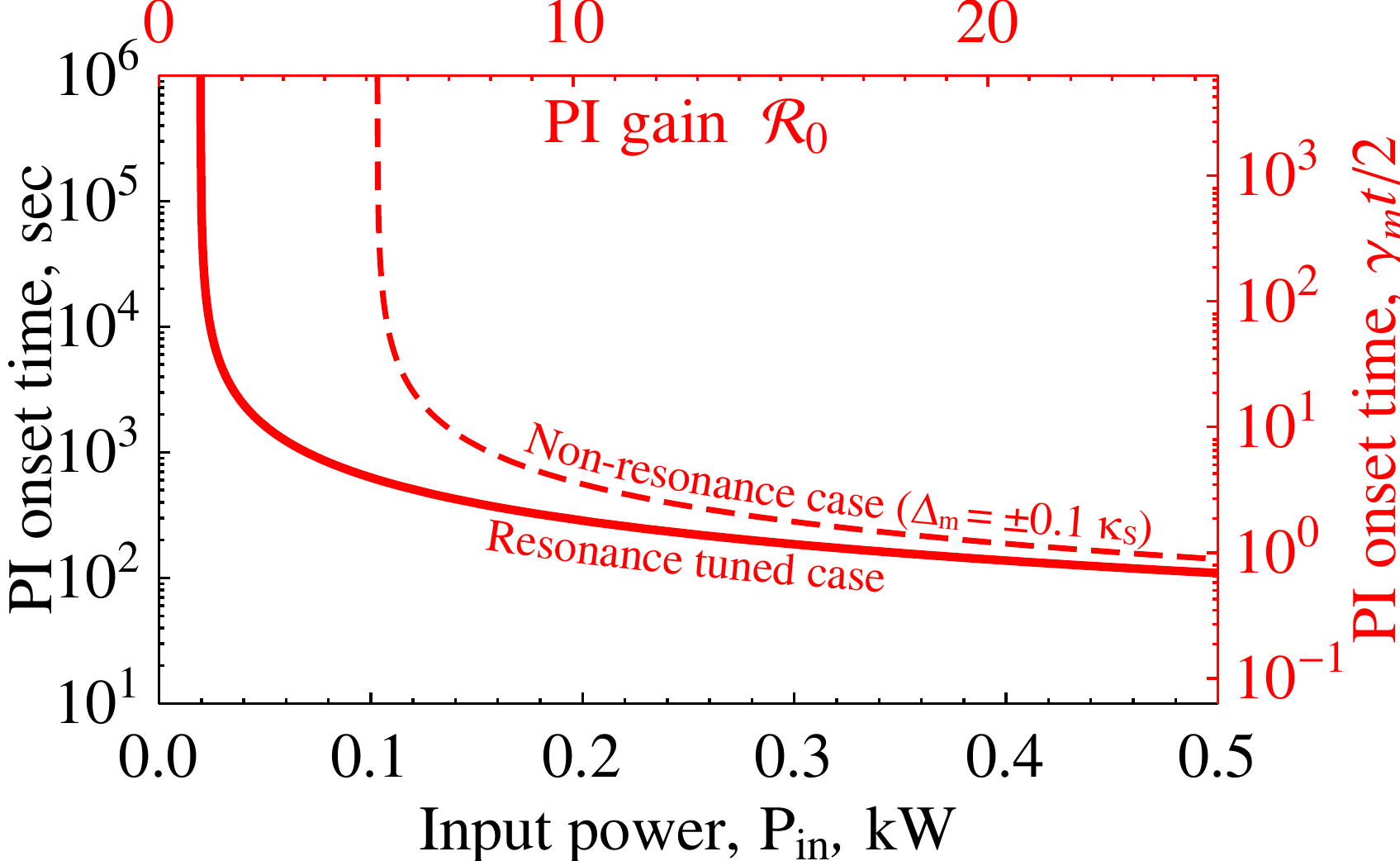}
\caption{Characteristic time for non-linearity in parametric instability to take over. Thin dashed line shows the effect of non-zero frequency mismatch ($\Delta_m = \pm 0.1\kappa_S$ for this plot).
 }\label{fig:PI_onset_time}
\end{figure}

The dependence of PI onset time vs. pump laser power $P_{in}$ (and PI gain) is plotted in Fig.~\ref{fig:PI_onset_time} for parameters characteristic for Advanced LIGO detector and given in Table~\ref{tab:params}. 

\section{Dual recycling interferometer}\label{sec:aLIGO_PI}

\begin{figure}[ht]
 \includegraphics[width=.49\textwidth]{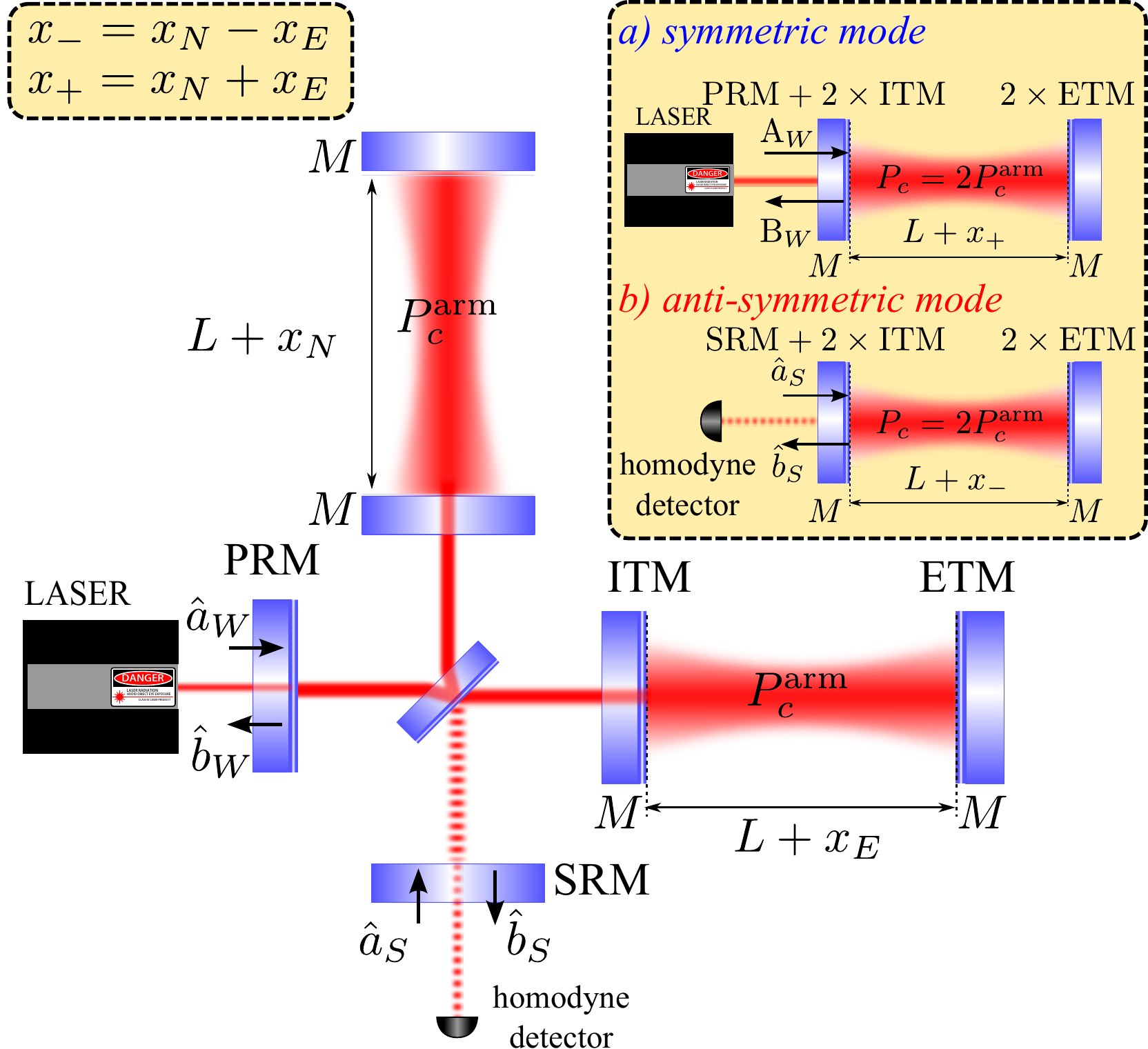}
\caption{Scaling law for dual recycled Advanced LIGO interferometers: \textit{common} and \textit{differential} modes of a balanced interferometer, representing sum and difference of the optical fields in the arm cavities, can be modelled as two independent effective Fabry-P\'erot cavities coupled to \textit{common} and \textit{differential} acoustic modes of the arm cavities mirrors. 
 }\label{fig:ScalLaw}
\end{figure}  

The above results are obtained for a single Fabry-P\'erot cavity. However, they are easily generalised to the case of a power and signal recycled interferometer like the one of Advanced LIGO. Such an interferometer, if perfectly symmetric, is equivalent to two effective Fabry-P\'erot interferometers with effective linewidths, $\kappa_{0\pm}$, and detunings, $\delta_\pm$, defined in terms of arm cavities parameters and power/signal recycling mirror reflectivity and phase shift. Graphically, this fact is illustrated in Fig.~\ref{fig:ScalLaw}. This result is well known as ``scaling law'' and devised by Chen and Buonanno in their seminal article \cite{2003_PhysRevD.67.062002_ScalLaw}. 

The effective Fabry-P\'erot cavities represent so-called \textit{symmetric} and \textit{anti-symmetric} optical modes, which are coupled to corresponding \textit{common} and \textit{differential} acoustic modes of the mirrors (see related definitions in Sec.~5.3 of \cite{Liv.Rv.Rel.15.2012}). As shown in \cite{2007_Phys.Lett.A.362.91_Gurkovsky}, the above consideration of 3-mode parametric instability in Fabry-P\'erot applies to each of these modes with the following substitutions of parameters:
\begin{align}
M&\to M\,,\\
P_c &\to 2 P_c^{\rm arm}\,,\\
\kappa_0 &\to\kappa_{0\pm} = \kappa_0\Re\left[\dfrac{1-\rho_{p, s}e^{2i\phi^0_{p,s}}}{1+\rho_{p,s}e^{2i\phi^0_{p,s}}}\right]\,,\\
\kappa_S &\to\kappa_{S\pm} = \kappa_S\Re\left[\dfrac{1-\rho_{p, s}e^{2i\phi^S_{p,s}}}{1+\rho^S_{p,s}e^{2i\phi_{p,s}}}\right]\,,\\
\Delta_m &\to\Delta_m + \delta_\pm = \Delta_m + \kappa_0\Im\left[\dfrac{1-\rho_{p, s}e^{2i\phi_{p,s}}}{1+\rho_{p,s}e^{2i\phi_{p,s}}}\right]\,,
\end{align}
where $\rho_{p,s} = \sqrt{1-T_{p,s}}$ are (amplitude) reflectivities of power and signal recycling mirrors, respectively, (with $T_{p,s}$ being more habitual power transmissivities thereof), and $\phi^{0,S}_{p,s} = \omega_{0,S} l_{p,s}/c$ are propagation phase shifts light of frequency $\omega_{0,S}$ acquires propagating the distance $l_{p,s}$ from arm cavities input test masses (ITMs) to the PRM and SRM, respectively. Here ``+'' sign stands for values that refer to symmetric mode, and ``$-$'' sign indicates those of anti-symmetric one. 

One can notice that the new linewidths ratio $\kappa_{0\pm}/\kappa_{S\pm}$ may be different from the one for the simple Fabry-P\'erot, as phase shifts $\phi^{0}_{p,r}$ deviate from $\phi^{S}_{p,r}$, \textit{i.e.} the fact that fundamental mode is resonant in PR/SR cavity does not mean that the same is true for the Stokes mode. The difference is $\Delta\phi = \phi^{0}_{p,r} - \phi^{S}_{p,r} \simeq \Omega_m l_{p,r}/c$ which might be significant for $\sim 25$ metres long PR/SR cavities planned for Advanced LIGO.

Asymmetry in interferometer arms does not change the general conclusion of this section, for the dual recycled interferometer can still be represented as two independent Fabry-P\'erot interferometers, as shown in \cite{2007_PLA.365.1.10_StriginPI,2010_CQG.27.205019_Gras_et_al}. In this case, however, normal modes of the asymmetric system are not pure \textit{symmetric} and \textit{anti-symmetric} combinations of fields in the arm cavities, but a general linear combination thereof  (cf. Section 2.6 of \cite{2007_PLA.365.1.10_StriginPI})

\section{Conclusion}

In this work, we analysed, using full non-linear treatment, the dynamics of 3-mode optomechanical instability in large-scale gravitational wave interferometers with freely suspended optics. It turns out that intrinsic non-linearity of the 3-mode interaction does not allow excited unstable optical and acoustic mode amplitudes to grow unbounded, rather it makes them saturate to the new steady state values. These values are governed by 3 dimensionless parameters: parametric gain, $\mathcal{R}_0$, normalised frequency mismatch (detuning), $\delta_m = 2(\omega_0-\omega_S-\Omega_m)/\kappa_S$, and the ratio of optical linewidths $\kappa_0/\kappa_S$. Therefore, our theory can be equally applied to simple Fabry-P\'erot interferometers and to complex dual recycled interferometers of the second generation gravitational wave detectors, such as Advanced LIGO \cite{Thorne2000,Fritschel2002}, Advanced Virgo \cite{Acernese2006-2}, KAGRA \cite{KAGRA_paper_Somiya} and GEO-HF\cite{Willke2006}.

Our analysis shows that the process of instability development is quite slow for large scale detectors, lasting for many relaxation times of the acoustic modes, which, due to very high Q factors of these modes, amounts to hundreds to thousands of seconds. Such a long onset time allows for efficient control and mitigation of this type of instability. Moreover, for reasonable values of PI gain, $\mathcal{R}_0\sim 10$, consistent with recent parametric instability modelling for Advanced LIGO interferometers \cite{2009_CQG_Gras_et_al,2010_PLA_Evans_et_al}, acoustic mode amplitudes should not exceed nanometre level. Basing on this point as well as on the fact that the small-scale membrane-in-the-middle interferometer in UWA where PI was observed experimentally in the non-linear regime \cite{2014_arXiv1411.3016_Sundae_paper} did not loose lock, one can presume that the same would be the case for a large-scale GW detector. However, there remains a high level of uncertainty pertaining to a greater complexity of an electronic control of a large interferometer. The response of the electronic control system to a slump of circulating power in the arms when PI starts to develop deserves a separate study. 
%
%that even if 3-mode instability starts develops to the stage when the acoustic amplitude reaches its steady state value, the interferometer is likely to remain locked, allowing to undertake necessary measures to suppress instability without relocking the whole system.

% leads to saturation of excited  amplitudes at the level governed by 
%
%We showed that contrary to linearised theory predictions, the initial exponential growth of acoustic and higher-order optical modes amplitudes is bound to stop after a certain period of time and saturate. The new steady state amplitude values for all three participating modes, as well as the instability development timescale are governed by 3 dimensionless parameters: parametric gain, $\mathcal{R}_0$, normalised frequency mismatch (detuning), $\delta_m = 2(\omega_0-\omega_S-\Omega_m)/\kappa_S$, and the ratio of optical linewidths $\kappa_0/\kappa_S$. It is therefore obvious that these results are universal and can be scaled to any scheme of gravitational wave interferometer using appropriate scaling laws \cite{2003_PhysRevD.67.062002_ScalLaw,Liv.Rv.Rel.15.2012}.

It should be also noted that our theory assumes fixed, time-independent values of mode frequencies and optomechanical coupling. In real interferometers with suspended optics mirrors tend to move around slowly and a laser beam spot does not rest at a fixed position on a mirror surface. This results in slow compared to acoustic frequencies modulation of the frequencies of the Stokes modes and thereby of the mismatch parameter, $\Delta_m(t)$, and of the PI gain $\mathcal{R}_0(t)$  and of modes overlap factor $\Lambda$, leading to reduced chance of instability. The effect of such modulation will be reported elsewhere \cite{paper_in_preparation}. 

\section{Acknowledgements.}

Authors are grateful to Andrey Matsko, Haixing Miao and Huan Yang for fruitful and elucidating discussions, as well as to all the members of the Macroscopic Quantum Mechanics discussion group for helpful comments and suggestions on improvement of our manuscript. We also thank anonymous referees for constructive critics and useful comments that, we hope, helped refine the presentation and quality of our manuscript substantially. The work of D.G.B, J.L. and C.Z. is supported by the Australian Research Council.

\end{document}